\documentclass[%
 reprint,
superscriptaddress,
 amsmath,amssymb,
 aps,
prl,
]{revtex4-2}

\usepackage{graphicx}
\usepackage{dcolumn}
\usepackage{bm}
\usepackage{hyperref}

\usepackage{physics}
\usepackage{ulem}
\usepackage[english]{babel}
\usepackage{soul}
\usepackage{todonotes}
\usepackage[english]{babel}
\usepackage{float}

\begin{document}

\preprint{APS/123-QED}

\title[0D Orbitronics]{Orbitronics in Zero-dimensional Graphene Nanoflakes via Electrode Coupling}

\author{Washington F. dos Santos}
\email{w.santos-junior@unesp.br}
\affiliation{Instituto de Física Teórica, S\~ao Paulo State University, S\~ao Paulo, Brasil}

\author{Eudes Gomes da Silva}
\affiliation{Department of Physics and Astronomy, University of Iowa, Iowa City, Iowa 52242, USA}

\author{Denis R. Candido}
\email{denis-candido@uiowa.edu}
\affiliation{Department of Physics and Astronomy, University of Iowa, Iowa City, Iowa 52242, USA}
\affiliation{Applied Mathematical \& Computational Sciences, The University of Iowa, Iowa City, Iowa, USA}

\author{Alexandre Reily Rocha}
\email{alexandre.reily@unesp.br}
\affiliation{Instituto de Física Teórica, S\~ao Paulo State University, S\~ao Paulo, Brasil}

\date{\today}

\begin{abstract}
We show that graphene nanoflakes realize a previously unexplored form of orbitronics in which orbital currents emerge in the absence of intrinsic angular momentum of the constituent atoms. Starting with a hexagonal graphene nanoflake, we demonstrate the presence of energy-degenerate eigenstates with opposite non-zero expectation values of the angular momentum. Placing electrodes in an appropriate configuration on the flake couples the reservoirs more strongly to one of the two opposite-angular-momentum states, yielding a chiral current that carries a net orbital angular momentum. The resulting circulating orbital currents generate magnetic fringe fields that are experimentally accessible with spin centers or superconducting quantum interference magnetometers, providing avenues for their experimental realization. Finally, our work establishes a new platform for exploring and studying orbitronics in zero-dimensional systems.
\end{abstract}

\maketitle



Orbitronics is the emerging field in solid-state physics that exploits the use of orbital angular momentum as a carrier of information \cite{Bernevig2005OrbitalHall,choi_observation_2023}, in close analogy to the role played by charge and spin in the well-established fields of electronics \cite{Dattamesoscopic} and spintronics \cite{PhysRevLett.61.2472,PhysRevB.39.4828,wolf2001spintronics}, respectively. Since its initial proposal \cite{Bernevig2005OrbitalHall}, several works — primarily focusing on three-dimensional systems \cite{Kontani2009GiantOHE,Go2018IntrinsicOHE} and, more recently, on two-dimensional materials \cite{He2020GiantOrbital,Cysne2021OrbitalHall2D} — have investigated mechanisms for generating orbital-polarized currents. The conventional picture of orbitronics is based on local atomic orbitals carrying nonzero angular momentum \cite{Cysne2021OrbitalHall2D,orbitalHalleffectTMD}. External electric fields and magnetic fields, provide access to these otherwise compensated orbital degrees of freedom,   by explicitly lifting orbital degeneracies \cite{Aivazian2015,PhysRevLett.121.086602}. 
 More recently, however, nonzero net orbital angular momentum (OAM) has been reported even in the absence of local atomic OAM contributions. Instead, it emerges through magnetic-field induced confinement \cite{Ge2023GiantOrbital}, localized edge states \cite{Busch2023,Lee2024}, engineered honeycomb lattices \cite{Bhowal2021,Cysne2024}, or intrinsic structural chirality \cite{Yoda2018,Gobel2025,Oh2026}.

\begin{figure}[htp!]
\centering
\includegraphics[width=0.5 \textwidth]{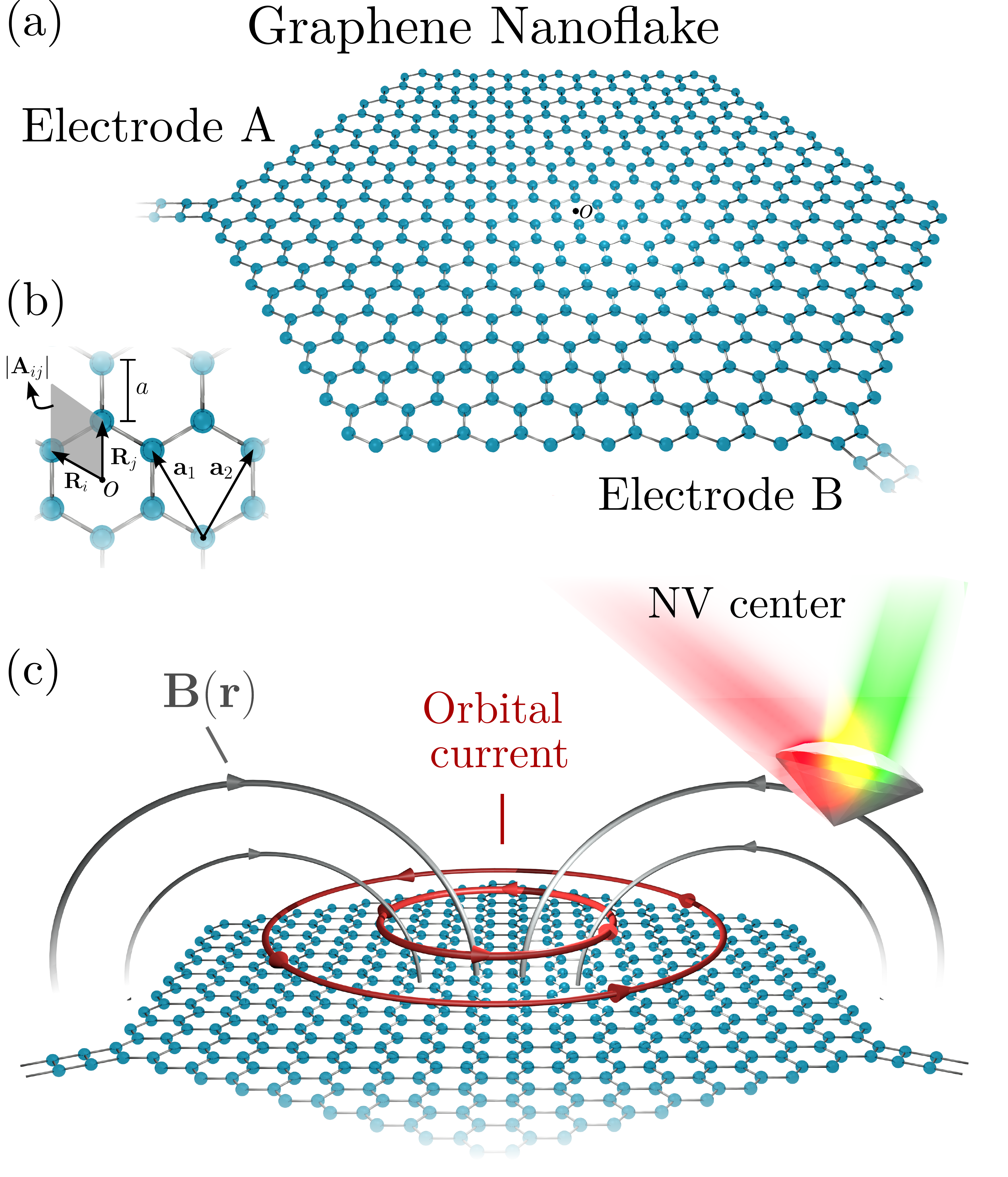}
\caption{(a) Hexagonal GNF connected to two electrodes in the asymmetric meta configuration. The leads consist of two semi-infinite chains of atoms connected to each other. (b) Schematic representation of the graphene lattice with Bravais lattice vectors ($\textbf{a}_1$ and $\textbf{a}_2$), and the oriented area $\textbf{A}_{ij}$ associated with site positions $\mathbf{R}_i$ and $\mathbf{R}_j$ defined from the origin. (c) Schematics of the generation of orbital current, associated magnetic field $\mathbf{B}(\mathbf{r})$, and a nitrogen-vacancy (NV) center for its experimental realization. The arrows indicate the direction of particle flow.}\label{estr}
\end{figure}

In this Letter, we demonstrate a different orbital paradigm. In contrast to standard descriptions of orbitronics, where OAM is inherited from local atomic orbitals, we report that a collective angular momentum can emerge in a zero-dimensional system coupled to leads [Fig.~\ref{estr}(a)], even when all atoms carry zero OAM, and there is no spin-orbit coupling. More specifically, we show that hexagonal graphene nanoflakes (GNF) \cite{nanoflakes,nanoflakesProp} described entirely by $p_z$ orbitals present degenerate energy eigenstates with finite and opposite orbital angular momentum. Leads positioned at the vertices of our flakes in the low-bias regime [See Fig.~\ref{estr}(a)] couple more strongly to one of these degenerate states. This imbalance favors a circulation direction, producing the formation of a net chiral circulating current (or current vortices)~\cite{current3,current,cir_cur_devices,levitov2016electron,palm2024observation,aharonsteinberg2022direct} carrying non-zero OAM, illustrated schematically in Fig.~\ref{estr}(c). Interestingly, our chiral currents are not restricted to the low-energy physics of graphene zigzag edge states; rather, these circulating currents can be observed over the entire structure, even in the presence of defects~\cite{current}. Finally, we calculate the magnetic fringe fields originating from our orbital currents and predict those can be detected via magnetometry techniques with nitrogen-vacancy centers (NV)~\cite {rondin2014magnetometry,schirhagl2014nitrogen,awschalom2018quantum,hong2013nanoscale,maze2008nanoscale} or 
Superconducting Quantum Interference Device (SQUID)~\cite{marchiori2022nanoscale,fagaly2006superconducting} [See Fig.~\ref{estr}(c)]. This introduces a novel approach to experimentally realize orbital currents, distinct from magneto-optical detection of accumulated orbital magnetic moment~\cite{lyalin2023magneto,choi_observation_2023} and electrical measurements detecting the orbital-charge conversion~\cite{hayashi2024observation,santos2026probing}. 

\begin{figure}[ht!]
\centering
\includegraphics[width=0.48 \textwidth]{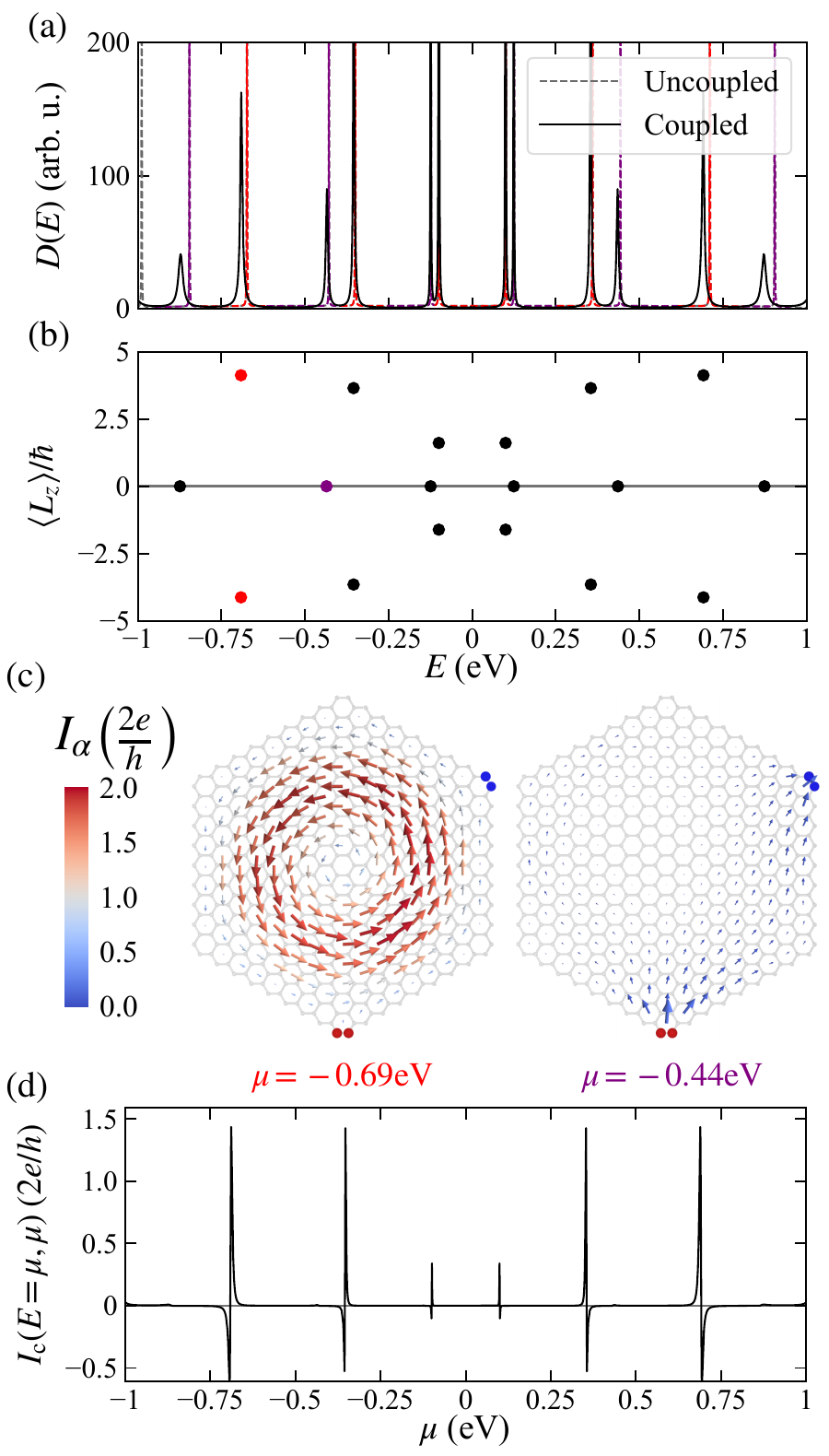}
\caption{(a) Density of states of the graphene nanoflake (GNF) with (black solid line) and without (colored dashed lines) coupling to the leads. (b) OAM expectation values of the isolated-GNF eigenstates, ordered by energy. (c) Local current densities at representative energies for which a circulating current is present (${\mu}={E}=-0.69$ eV) and absent (${\mu}={E}=-0.44$ eV), respectively, with $V_b =0.02$ eV, and the arrows indicate the direction of particle (electron) flow by convention.  Red and purple indicate representative doubly-  and non-degenerate states, respectively. (d) Degree of circular current [Eq.~\ref{circularcurrent}] as a function of $\mu$ for $E=\mu$.}
\label{dos_L_transm_ccurr}
\end{figure}

Our system consists of a planar, hexagonal GNF comprising 486 carbon atoms, with zigzag edges joined by vertices with armchair termination [Fig.~\ref{estr}(a)]. Assuming the origin at the center of our nanoflake, site $i$ is ascribed a position vector, $\mathbf{R}_{i} = \left(X_{i},Y_{i}\right) = m_{i} \mathbf{a}_1 + n_{i} \mathbf{a}_2 + l_{i} a\mathbf{e}_x$, with $\left\{m_{i}, n_{i}\right\} \in \mathbb{Z},~l_{i} = \left\{1,2\right\}$, $\mathbf{a}_{1,2}$ the lattice vectors and $a$ the distance between carbon atoms [See Fig.~\ref{estr}(b)]. Here, $m_i$, $n_i$ and $l_i$ are appropriately chosen to reconstruct our hexagonal GNF. The carbon atoms are described by a single $p_z$ orbital for which the out-of-plane component of the
orbital angular momentum, $\hat{L}_z$, is zero. Thus, this establishes a minimal model that ensures that no atomic OAM is initially present. Within our tight-binding approximation, the Hamiltonian in second quantization reads~\cite{wallace1947band,reich2002tight,castro2009electronic}
\begin{eqnarray}
 \hat{\mathcal{H}}^{GNF} & = & \sum_{i} \epsilon_{0}\, d^\dagger_{i} d_{i}
 + \sum_{\langle i,j \rangle} t_{ij}\, d^\dagger_{i} d_{j} + h.c.,
\end{eqnarray}
where $\epsilon_{0} = 0 ~ eV$ is the on-site energy, and $t_{ij} = t =-3.1 ~ eV$ \cite{reich2002tight} is the nearest-neighbor hopping parameter between atomic sites $i$ and $j$. The operator $d^\dagger_{i}$ ($d_{i}$) is the electron creation (annihilation) operator associated with the $p_{z}$ orbital at site $i$ of the nanoflake, \textit{i.e.}, $d_i^\dagger |0\rangle=|i\rangle$, with $\langle \textbf{r}|i\rangle = \phi_{p_z}(\textbf{r}-\textbf{R}_i)$ and $|0\rangle$ the vacuum.  Diagonalization of this Hamiltonian yields eigenvectors $\{ |\psi^{\epsilon_n} \rangle\}$ and corresponding energies $\{\epsilon_{n}\}$ that are plotted as dashed vertical lines in Fig.~\ref{dos_L_transm_ccurr}(a), with purple (red) dashed lines representing single (double) degenerate states. We now turn our attention to characterizing these degeneracies. For that, we calculate the expectation value of $\hat{L}_z$ with respect to the eigenstates of our Hamiltonian and plot them in Fig.~\ref{dos_L_transm_ccurr}(b). For the non-degenerate states, we find $\langle \hat{L}_z \rangle=0$ [see correspondence between purple lines in Fig.~\ref{dos_L_transm_ccurr}(a) and dots in Fig.~\ref{dos_L_transm_ccurr}(b)]. 
Conversely, projecting $\hat{L}_z$ onto a set of degenerate eigenstates and diagonalizing it results in opposite expectation values for $\hat{L}_z$ [see correspondence between red lines in Fig.~\ref{dos_L_transm_ccurr}(a) and dots in Fig. \ref{dos_L_transm_ccurr}(b)].

Henceforth, we choose the doubly-degenerate $n$-eigenstates with energy $\epsilon_n$ such that they have opposite expectation values of $\hat{L}_z$, \textit{i.e.}, $\{|\psi_{\langle L_z\rangle>0}^{\epsilon_n}\rangle, |\psi_{\langle L_z\rangle <0}^{\epsilon_n}\rangle\}$. Our results demonstrate that degenerate eigenstates carry opposite {\it non-zero orbital angular momentum} despite the vanishing atomic OAM contribution from the $p_z$ orbitals. To understand the underlying origin of this, we project $\hat{L}_z$ onto the basis of tight-binding orbitals  $(|i\rangle)$, that correspond to solutions of the isolated atoms obtained via, e.g., density functional theory methods~\cite{dft1}. Assuming an overlap matrix between adjacent sites $\mathcal{S}_{ij}=\langle i|j\rangle\ll 1$, we obtain (See End Matter)
\begin{equation}
    \langle i|L_z|j\rangle = \frac{i m_e}{\hbar}t_{ij} \textbf{A}_{ij}, \label{lz-area}
\end{equation}
where $m_e$ is the electron bare mass, $\hbar$ the reduced Planck constant and $\textbf{A}_{ij}=(\textbf{R}_i \times \textbf{R}_j)_z$ is the oriented rhombus area with sides $\textbf{R}_i$ and $\textbf{R}_j$, shown in Fig.~\ref{estr}(b). This demonstrates that OAM can emerge in our system even in the absence of its atomic counterpart. This is similar to earlier works that demonstrated non-zero angular momentum arising from the linear combination of multi-bands with zero OAM~\cite{Busch2023,Lee2024,Bhowal2021,Cysne2024,Yoda2018,Gobel2025,Oh2026}. For a more general case, in which the atomic contribution is present, the total orbital angular momentum is a combination of the intrinsic atomic part and the geometric part obtained here (see End Matter for the full multi-orbital non-orthogonal derivation
and Supplemental Material~\cite{SM} for additional details).

\begin{figure*}[t!]
\centering
\includegraphics[width=\textwidth]{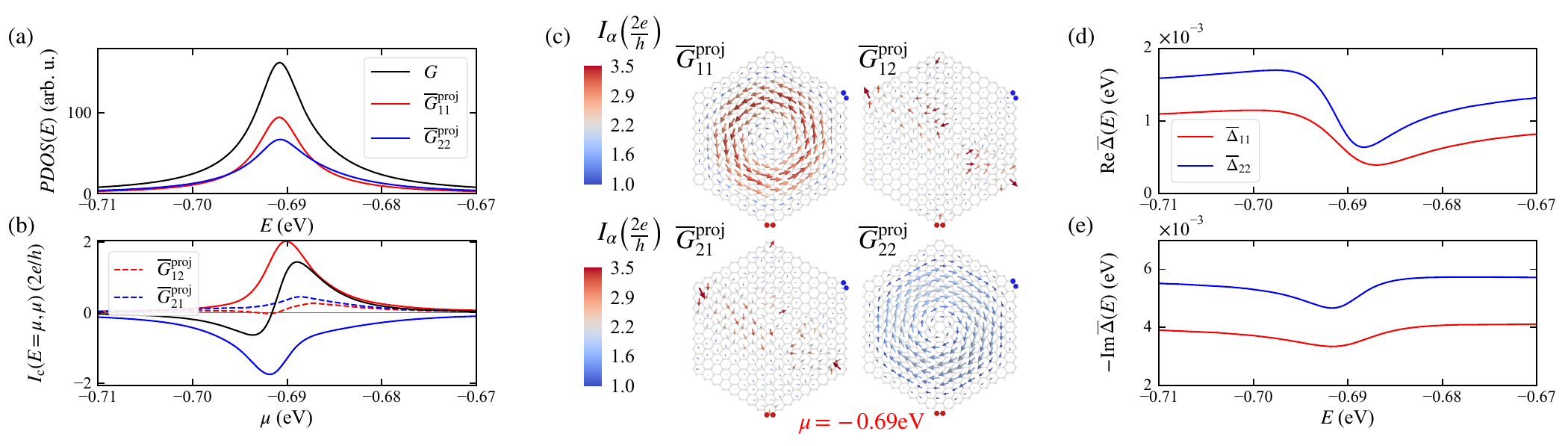}
\caption{(a) Total density of states of the GNF (solid black line), and corresponding projections $\overline{G}^{\textrm{proj}}_{11}$ (solid red line) and $\overline{G}^{\textrm{proj}}_{22}$ (solid blue line) onto $\psi^{1}_{\langle \hat{L}_z\rangle}$ and $\psi^{2}_{\langle \hat{L}_z\rangle}$ with energy $\varepsilon_E =-0.69 ~ eV$. (b) Total circular current (black solid line) and its components considering only selected matrix elements of the projected Green's function as non-zero. Solid lines correspond to diagonal components, and dashed lines to off-diagonal ones. (c) Local current density considering each matrix element $\overline{G}^{\textrm{proj}}_{11}$, $\overline{G}^{\textrm{proj}}_{12}$, $\overline{G}^{\textrm{proj}}_{21}$, and $\overline{G}^{\textrm{proj}}_{22}$ of the projected retarded Green's function expressed in the $\hat{L}_z$ eigenbasis. (d) Real and (e) negative imaginary parts of each diagonal component of the hybridization function [Eq.~(\ref{hybri})].}\label{dos-0.68}
\end{figure*}

Despite the interesting emergence of states carrying non-zero OAM in our nanoflake, the {\it total} angular momentum is always zero for a fixed chemical potential, thus precluding the generation of a net OAM. As a consequence, the manifestation of a net OAM in this system requires either lifting the degeneracy~\cite{Ge2023GiantOrbital} or having transport emerging dominantly from one of the degenerate states with opposite angular momentum. We accomplish the latter by coupling our GNF to two semi-infinite charge reservoirs attached to the hexagonal vertices as shown in Fig.~\ref{estr}(a). Each lead {(A and B)} consists of two lines of atoms {that are also vertically} coupled to each other. For simplicity, the leads are assumed to be made of graphene. With this, our total Hamiltonian becomes ${\cal \hat{H}}={\cal \hat{H}}^{GNF}+{\cal \hat{H}_{A}}+{\cal \hat{H}_{B}}+{\cal \hat{V}}^{GNF}_{A}+{\cal \hat{V}}^{GNF}_{B}$ with
\begin{eqnarray}
 \hat{\mathcal{H}}_{A(B)} & = & 
  \sum_{k=0}^{\infty} \sum_{l=1,2}
 \bar{t}_{k,  k+1}^{l, l}
 c^{A(B)\dagger}_{l,k}\, c^{A(B)}_{l,k+1} \nonumber \\
 & + & \sum_{k=0}^{\infty} {\bar{t}}_{k,k}^{1,2}\,
 c^{A(B)\dagger}_{1,k}\, c^{A(B)}_{2,k}
 + h.c., \\
 \hat{\mathcal{V}}^{GNF}_{A(B)} & = &
 \sum_{i \in \{ \mathrm{edge}\, A(B) \}}
 \sum_{l=1,2}
 \gamma_{l}^{A(B)}\, 
 d^\dagger_{i}\, c^{A(B)}_{l,k=0}
 + h.c.,
\end{eqnarray}
where $c_{l,k}$ ($c_{l,k}^\dagger$) represent the creation (annihilation) operator associated with the lead atoms, $k=0,..,\infty$ the length index associated with the semi-infinite direction, and $l=1,2$ the width index, $\overline{t}_{k,k'}^{l,l'}  = t = -3.1$ eV, and $\gamma^{A(B)}_l = 1$ eV. Following the formalism of Caroli \textit{et al.}.~\cite{Caroli_1971}, we write the retarded Green's function for the scattering region—represented by the GNF—tracing out the effects of the reservoirs \cite{Dattamesoscopic},
\begin{equation}
    G^{r}\left(E\right)
    =
    \lim_{\eta \to 0^+}
    \left[
    (E+i\eta)I
    -
    \hat{\cal{H}}^{GNF}
    -
    \Sigma^{A}\left(E\right)
    -
    \Sigma^{B}\left(E\right)
    \right]^{-1},
    \label{Gret}
\end{equation}
where $\Sigma^{A}$ ($\Sigma^{B}$) is the self-energy associated with lead A (B), and is computed recursively using Dyson's equation \cite{Rungger2008}. The density of states (DOS) projected onto atomic orbitals of the full system is obtained via 
\begin{equation}\label{D}
D(E) = -\frac{1}{\pi} \Tr \big[ Im \{G^{r}\left(E\right)\} \big],
\end{equation}
where we trace with respect to the atomic orbitals. In Fig.~\ref{dos_L_transm_ccurr}(a) we plot the DOS as solid black lines, where the coupling of our flake to the leads broadens (shifts) the flake's energy levels via the imaginary (real) part of the self-energy. Despite the strong coupling strength between atoms of the GNF and leads ($\sim 1$~eV), our system is demonstrated to be in the weak-coupling regime (\textit{i.e.}, small broadening compared to the energy separation between energy levels). This happens because the electrode--flake contact is restricted to only a small number of carbon atoms at each of the GNF vertices.

The finite bias voltage applied between leads ensures the flow of a charge current~\cite{Dattamesoscopic}. The spatial distribution of these currents is calculated using the non-equilibrium Keldysh's formalism~\cite{Caroli_1971,dos_Santos_2025}, with current density from site $i$ to $j$ along the bond direction $\mathbf{e}_{ij}=\left(\mathbf{R}_j-\mathbf{R}_i\right)/\left| \mathbf{R}_j-\mathbf{R}_i \right|$~\cite{cresti2003keldysh} given by
\begin{equation}\label{locadenscurrent} 
    {\mathbf{I}_{i \rightarrow j}}\left(E,\mu\right) = \frac{2e}{h}  \textrm{Re} \left\{ \hat{\cal{H}}^{GNF}_{ij}G_{ji}^{<}\right\}\mathbf{e}_{ij},
\end{equation} 
with, $e$ the magnitude of the  electron charge, $h$ Planck's constant, ${\cal \hat{H}}_{ij}^{GNF}=\langle i|{\cal \hat{H}}^{GNF} | j \rangle$ and the matrix elements of the lesser Green's function $G_{ij}^{<}=\langle i|G^{<} | j \rangle$. For a non-interacting system, $G^{<}\left(E,\mu,V_b\right) = i G^{r}\left(E\right)\left[\Gamma^A\left(E,V_b\right) f_A\left(E\right) + \Gamma^B\left(E,V_b\right)f_B\left(E\right)\right] G^{a}\left(E\right)$~\cite{Dattamesoscopic}, $\Gamma^{A(B)} = i[\Sigma^{A(B)} - (\Sigma^{A(B)})^{\dagger}]$ is the imaginary part of the self-energy, and time-reversal symmetry ensures  $G^r = \left(G^a\right)^\dagger$. 
$f_{A(B)}\left(E\right)=f\left(E,\mu_{A(B)}\right)$ is the Fermi-Dirac distribution of lead A (B), with chemical potential $\mu_A=\mu+V_b/2$ ($\mu_B=\mu-V_b/2$). The equilibrium chemical potential $\mu$ is controlled via a backgate~\cite{bischoff2015localized} and the external bias potential is set to $V_b =0.02$~eV to keep our system in the low-bias regime. 

To spatially represent the local current density, we compute the current at the center of each hexagonal ring $\alpha$ via ${\mathbf{I}_{\alpha}(E,\mu)} ={\sum}_{\left\{i,j\right\} \in \alpha} {\mathbf{I}_{i \rightarrow}j}\left(E,\mu\right)$. Fig.~\ref{dos_L_transm_ccurr}(c) shows ${\mathbf{I}_{\alpha}(E,\mu)}$ for representative chemical potentials $\mu={E}=-0.69$~eV and $\mu={E}=-0.44$~eV. Interestingly, for the former, we observe a chiral circulating current associated with the degenerate state with $\langle \hat{L}_z\rangle>0$, due to the counterclockwise circulation of the current. As a consequence, our circulating current is OAM-polarized and thus carries OAM. The right panel of Fig.~\ref{dos_L_transm_ccurr}(c) shows the non-circulating character of the current for the non-degenerate level with $\mu={E}=-0.44$~eV. Most importantly, the emergence of circulating current does not require fine-tuning of the parameters, and it is shown to be robust against differences in both the values of lead-flake coupling  $\gamma_l$ and the positions of the lead (See Supplemental Material~\cite{SM}).  Nevertheless, the circulating current can also vanish, for instance, when the leads are attached to vertices along the vertical direction (See Supplemental Material~\cite{SM}). Circulating currents (or current vortices) have been reported in both molecules \cite{cir_cur_devices,current2,current3,current4}, graphene-based structure~\cite{current,current4}, spin defects in quantum wells \cite{da2023dissipationless} and quantum dots~\cite{van2014geometric,van2014spin,candido2018blurring}, and have been shown to be robust even in the presence of interactions \cite{vortox_ee_phono}, defects and under high bias voltages~\cite{current}. 

To understand the origin and requirements for the emergence of a circulating current, we extend the measure described in Refs.~\cite{current3,cir_cur_devices} to our system (See Supplemental Material~\cite{SM}) and calculate the degree of circulating current associated with our states using 
\begin{equation}\label{circularcurrent} 
I_{c}\left(E, \mu \right) = \sum_{\alpha}^{N_{\alpha}}\frac{1}{l_{\alpha}}\left|\hat{\mathbf{I}}_{\alpha}\left(E, \mu \right)\right|{\rm sign}(\mathbf{R}_{\alpha}\times\mathbf{I}_{\alpha}),
\end{equation}
where $l_\alpha$ is the perimeter of individual hexagonal ring $\alpha$ centered at $\mathbf{R}_\alpha$. Fig.~\ref{dos_L_transm_ccurr}(d) shows $I_c \left(E, \mu \right)$ for $\mu=E$ as a function of the chemical potential, and demonstrates that circulating currents only emerge in the energy windows containing doubly degenerate states. This hints that the origin of the circulating current is associated with how the leads couple to the degenerate eigenstates with opposite expectation value of $\hat{L}_z$. To investigate this mechanism further,  we adopt an approach~\cite{projec} consisted of projecting the total Green's function [Eq.~(\ref{Gret})] onto the subspace spanned by the degenerate eigenstates with $\epsilon_n \equiv -0.69$~eV, ${\{|\psi^{1}_{\langle \hat{L}_z\rangle}\rangle,|\psi^{2}_{\langle \hat{L}_z\rangle}\rangle \} \equiv \{|\psi^{\epsilon_n}_{\langle \hat{L}_z\rangle>0}\rangle,|\psi^{\epsilon_n}_{\langle \hat{L}_z\rangle<0}\rangle\}}$. This yields the projected Green's function
\begin{equation}
\overline{G}^{\mathrm{proj}} \left(E\right) =
 U_{\psi}^{\dagger} ~G^{r}\left(E\right) ~U_{\psi} ~,
\end{equation}
with $U_{\psi} = \left(|\psi^{1}_{\langle \hat{L}_z\rangle}\rangle,|\psi^{2}_{\langle \hat{L}_z\rangle}\rangle\right)^T$. We rewrite $G^r$ by performing the inverse projection, yielding
\begin{equation}\label{Grinv}
G^{r} \left(E\right) = U_{\psi} \left( \sum_{i,j} \overline{G}^{\mathrm{proj}}_{ij}\left(E\right) \, \ket{\psi^{\epsilon_i}_{\langle \hat{L}_z\rangle}}\bra{\psi^{\epsilon_{j}}_{\langle \hat{L}_z\rangle}} \right) U_{\psi}^{\dagger} ,
\end{equation}
where the diagonal elements represent the direct contributions from individual eigenstates, while the off-diagonal elements capture eigenstate mixing.

Since quantities such as the local current density [Eq.~(\ref{locadenscurrent})] and the degree of circular current [Eq.~(\ref{circularcurrent})] depend directly on $G^r$, Eq.~(\ref{Grinv}) allows us to identify the contribution arising from the different components of $\overline{G}_{ij}^{\textrm{proj}}$. Fig.~\ref{dos-0.68}(a) shows the projected density of states onto the $|\psi^{1}_{\langle \hat{L}_z\rangle}\rangle$ and $|\psi^{2}_{\langle \hat{L}_z\rangle}\rangle$ subspace, {$PDOS_i \left(E\right)=-1/\pi\Im{\overline{G}^\mathrm{proj}_{ii}}$} with $i=1$ and $2$, respectively. We note that both curves are similar to each other, differing only by a slight asymmetry in their broadening and stronger spectral weight for $|\psi^{1}_{\langle \hat{L}_z\rangle}\rangle$. Fig.~\ref{dos-0.68}(b) shows that $|\psi^{1}_{\langle \hat{L}_z\rangle}\rangle$ and $|\psi^{2}_{\langle \hat{L}_z\rangle}\rangle$ contribute with opposite circular-current polarizations, with a small contribution coming from the off-diagonal terms. Owing to their unequal spectral weights, the net current arising around {$\mu={E}=-0.69$~eV} does not vanish completely, giving rise to the finite net circulating current shown in Fig.~\ref{dos_L_transm_ccurr}(c). This is shown explicitly in Fig.~\ref{dos-0.68}(c), where we plot the corresponding spatial current density associated with the different components $\overline{G}_{ij}^{\textrm{proj}}$, demonstrating the larger amplitude of the circulating current associated with state $|\psi^{1}_{\langle \hat{L}_z\rangle}\rangle$.

To understand the reason for this, we calculate the effective coupling between the $\{|\psi^{1}_{\langle \hat{L}_z\rangle}\rangle,|\psi^{2}_{\langle \hat{L}_z\rangle}\rangle \}$  subspace and the electronic reservoirs via the so-called hybridization function  \cite{Jacob2010,dos_Santos_2025}
\begin{equation}
\overline{\Delta}\left({E}\right)
=
\left[
\left(E-
\epsilon_n\right) \hat{I}_{2\times 2}
-\left(\overline G^{\textrm{proj}}\left(E\right)\right)^{-1}
\right].
\label{hybri}
\end{equation}
In Figs. \ref{dos-0.68}(d)-(e), we present the real and imaginary parts of the diagonal elements of $\overline{\Delta}$ as a function of energy. Similar to the self-energies ($\Sigma^{A(B)}$), the real part of $\overline{\Delta}$ causes energy shifts, whereas its imaginary counterpart broadens the energies. In Fig.~\ref{dos-0.68}(e), the larger imaginary values of $\overline{\Delta}_{22}$ over $\overline{\Delta}_{11}$ demonstrates that $|\psi^{2}_{\langle \hat{L}_z\rangle}\rangle$ is coupled more strongly to the leads. This is the reason for the higher spectral weight of $|\psi^{1}_{\langle \hat{L}_z\rangle}\rangle$ compared to  $|\psi^{2}_{\langle \hat{L}_z\rangle}\rangle$ in Fig.~\ref{dos-0.68}(a). This lead-induced coupling imbalance is sufficient to break the exact cancellation between the two states, leading to a purely electronic way of generating circulating currents with OAM. This demonstrates that the net circulating current originates from the unequal hybridization of the two states with opposite expectation value of OAM.

Current methods for detection of orbitronics-related phenomena rely on the magneto-optical Kerr effect~\cite{Manchon2019SOTReview}, where the accumulation of angular momentum --- and hence magnetic moment --- modifies the polarization and intensity of the incident light. While this technique has been used for realizing the orbital Hall effect and orbital currents~\cite{choi_observation_2023}, they are limited in spatial resolution, which is set by the diffraction limit ($\lambda\sim 700-800$~nm). Due to the small spatial dimension of our flakes ($\ll \lambda$), alternative techniques are required. Here, we propose a new approach to realize circular orbital currents based on sensing their magnetic stray field with nanoscale scanning probes, e.g., NV  centers~\cite{rondin2014magnetometry,schirhagl2014nitrogen,awschalom2018quantum,hong2013nanoscale,maze2008nanoscale} or nano Superconducting Quantum Interference Device (nano-SQUID)~\cite{marchiori2022nanoscale,fagaly2006superconducting}.

The magnetic fringe field at position $\mathbf{r}$ due to the circulating current that originates from the local current density ${\mathbf{I}_{i\rightarrow j}(E,\mu)}$ [Eq.~(\ref{locadenscurrent})] is calculated via Biot-Savart law (See Supplemental Material~\cite{SM})
\begin{equation}
     \mathbf{B}(\mathbf{r,\mu})
    =
    \sum_{\langle i,j\rangle}
    \int_{\mu-\frac{V_b}{2}}^{\mu+\frac{V_b}{2}}
    \frac{\mu_0 {\mathbf{I}_{i\rightarrow j}(E,\mu)}}{4\pi}\,\dd E
    \int_0^{d_{ij}}
    \frac{
        \dd\boldsymbol{\ell}_{ij}
        \times
        \mathbf{r}_{ij}'(\mathbf{r},l)
    }{
        \left|
            \mathbf{r}_{ij}'(\mathbf{r},l)
        \right|^3
    } ~,
     \label{eq:biot_savart}
\end{equation}
where $\mathbf{r}_{ij}'(\mathbf{r},l)
=
\mathbf{r}-(\mathbf{R}_i+l\mathbf{e}_{ij})$, $\boldsymbol{l}=l\mathbf{e}_{ij}$ is a vector along the bond between sites $i$ and $j$, and $d_{ij}=|\mathbf{R}_j-\mathbf{R}_i|$.

\begin{figure}[h]
\centering
\includegraphics[width=0.48\textwidth]{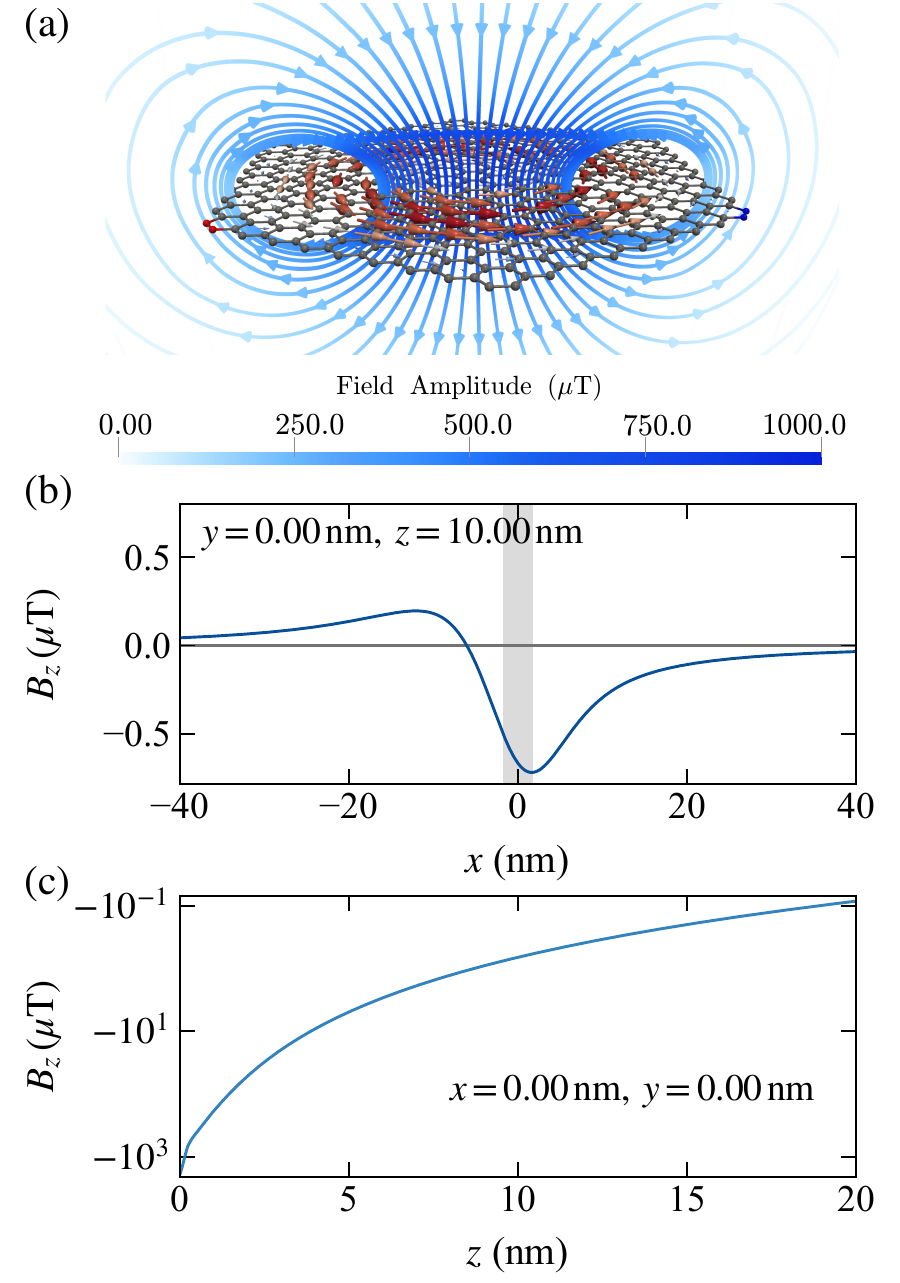}
\caption{(a) Magnetic fringe field $\mathbf{B}(\mathbf{r},\mu)$ due to the circulating orbital current of Fig.~\ref{dos_L_transm_ccurr}(c), \textit{i.e.}, $\mu=-0.69$~eV. The color scale represents the magnetic-field magnitude, while the arrows indicate its local direction. The red arrows indicate the intensity and direction of particle (electron) flow by convention. (b) Magnetic field $B_z$ as a function of $x$ at $y=0$~nm and $z=10$~nm. Shaded region represents the region occupied by the GNF. (c) Magnetic field $B_z$ as a function of $z$ at the center of our flake ($x=y=0$~nm).}\label{magfield}

\end{figure}

In Fig.~\ref{magfield}(a) we plot the magnetic fringe fields for the circulating current of Fig.~\ref{dos_L_transm_ccurr}(c) with $\mu=E = -0.69$~eV. The $z$-component of the magnetic field ($B_z $) is larger at the center of the nanoflake and decreases with increasing both radial distance and height, shown in Figs.~\ref{magfield}(b) ($y=0$, $z=10$ nm) and (c) ($x=y=0$), respectively. The magnetic field strength is always stronger than $0.1\ \mu$T over a spatial region of $\sim 20$~nm, which is greater than the spatial resolution of NV scanning probes ($\sim 10$~nm)~\cite{xu2025minimizing} and nano-SQUID ($\sim10-20$~nm)~\cite{vasyukov2013scanning,yang2026real}. Furthermore, typical sensitivities of NV centers in nanodiamond scanning tips are $\sim 1$~nT$/\sqrt{\textrm{Hz}}$ ~\cite{fang2013high,zhang2020high}, which, for an NV coherence time of $100\ \mu$s, yields a minimal detectable field $\delta B \sim 0.1\ \mu$T, lower than the values we report for the GNF. As a consequence, this places nanoscale scanning probe magnetometry as a promising avenue for realizing orbitronics effects in zero-dimensional systems.

In summary, we demonstrate that hexagonal graphene nanoflakes comprised of only $p_z$ orbitals develop doubly-degenerate eigenstates carrying opposite non-zero angular momentum, originating from the geometry of the system. Appropriately placing leads on our flake results in an unbalanced coupling to these states, yielding a net circulating orbital current with non-zero angular momentum. This circulating current is robust against different lead-flake coupling strengths. Finally, we propose a novel way for probing orbitronics in a zero-dimensional system based on nanoscale scanning probe magnetometers.

\textit{Acknowledgments}--- We thank Anderson L. R. Barbosa for fruitful discussions and acknowledge support from the Fundação de Amparo à Ciência e Tecnologia do Estado de Pernambuco (FACEPE), the National Council for Scientific and Technological Development (CNPq), and the Coordenação de Aperfeiçoamento de Pessoal de Nível Superior -- Brasil (CAPES; Finance Code 001). A.R.R. and W.F.S acknowledge support from the São Paulo Research Foundation (FAPESP; Grant Nos. 2026/03311-7, 2025/25217-0, 2025/24203-5, 2023/09820-2, and 2021/14375-0). Transport formalism, code development, and magnetometer analyses were supported by the National Science Foundation under Award No. ECCS-EPSRC2624415. Calculations were performed using CENAPAD-SP resources and the SDumont supercomputer at the National Laboratory for Scientific Computing (LNCC/MCTI, Brazil).

\bibliography{ref}

\onecolumngrid

\appendix
\setcounter{equation}{0}
\renewcommand{\theequation}{A\arabic{equation}}

\begin{center}
\textbf{End Matter}
\end{center}

\twocolumngrid

\textit{Matrix elements of the momentum and angular momentum operators}---
The Hamiltonian for the graphene nanoflake considered throughout is based on a one-electron Hamiltonian for a lattice of the form 
\begin{equation}
\hat{\mathcal{H}} = \frac{\mathbf{\hat{p}}^2}{2m_e} + \sum_i V\left(\mathbf{\hat{r}}-\mathbf{R}_i\right) ~,
\end{equation}
where $V$ is an effective single-particle potential representing an isolated atom at an arbitrary site $i$ with position $\mathbf{R}_i$ relative to the center of the GNF. In that sense, apart from including many-body interactions within a mean-field picture, we are starting from the atomistic description, not an effective one. Thus, in general, within the Heisenberg picture, the velocity operator for the whole system is given by
\begin{equation}
    \mathbf{\hat{v}} = \frac{i}{\hbar}\left[\mathcal{H},\mathbf{\hat{r}}\right] = \frac{1}{m_e}\mathbf{\hat{p}} . \label{eq:commutator}
\end{equation}
We now consider a basis consisting of the solutions of the isolated carbon atoms on each site  $|i,\alpha\rangle$, with $\alpha =\left\{n,l,m\right\}$ corresponding to the standard notation to describe the electronic configuration (of bound states) in atoms.
These orbitals can be obtained from, say, density functional theory \cite{dft1}. Assuming a complete, orthogonal basis, the projection on that basis is given by
\begin{equation}
\mathbf{\hat{p}}_{i\beta,j\beta^\prime} = m_e\mathbf{\hat{v}}_{i\beta,j\beta^\prime} ~.
\end{equation}
If we assume, as an approximation, that the eigenstates of the Hamiltonian that are of interest to us, can be written as a combination of $p_z$ orbitals, then that subspace is decoupled from the remaining atomic orbitals, which justifies one of the assumptions used in the main manuscript.

In general we can define the $z$-projection of the orbital angular momentum operator (OAM) in its symmetrized form as
\begin{eqnarray}
    \hat{L}_z & = & \frac{1}{2}\left[ \hat{\mathbf{r}} \times \hat{\mathbf{p}} - \hat{\mathbf{p}} \times \hat{\mathbf{r}}\right]_z, \nonumber\\
              & = & \frac{1}{2} \left( \hat{x}\hat{p}_y-\hat{y}\hat{p}_x - \hat{p}_x\hat{y}+ \hat{p}_y\hat{x}\right).\\
              & = & \frac{i m_e}{2\hbar} \left( \hat{x}\left[{\cal H},\hat{y}\right]-\hat{y}\left[{\cal H},\hat{x}\right] -  \left[{\cal H},\hat{x}\right]\hat{y} + \left[{\cal H},\hat{y}\right]\hat{x} \right).\nonumber 
\end{eqnarray}

In the main manuscript, we assume a subspace spanned by a subset of orbitals with index $\beta$ on each site. In our case, we consider only the $\beta = p_z$ orbitals (a single orbital per site). Moreover, the overlap between states in each site $\langle i | j \rangle = S_{ij}$ follows,
\begin{eqnarray}
 \mathcal{S}_{ii} & \gg & \mathcal{S}_{ij} ~, ~ i \neq j ~,\\ 
 \mathcal{S}_{ij} & \sim & 0 ~, ~ i \neq j ~,
\end{eqnarray}
and we assume no dipole terms,
\begin{equation}
   \langle i | \hat{\mathbf{r}} | j\rangle   =  \mathbf{R}_{i}\delta_{ij} ~. 
\end{equation}

As a consequence, we write  $\left[\hat{x},\hat{y}\right]=\hat{x}\hat{y}-\hat{y}\hat{x}=0$ in this basis. Thus, the equation for the angular momentum can be simplified to 
\begin{equation}
\hat{L}_z = \frac{ i m_e}{\hbar} \left(  \hat{x}{\cal H}\hat{y}-  \hat{y}{\cal H}\hat{x} \right) = \hat{x}\hat{p}_y-\hat{y}\hat{p}_x, \label{AOM}
\end{equation}
recovering the standard definition.

More importantly, within this orthogonal approximation, the commutation relation of Eq.~\ref{eq:commutator} still holds element-by-element within the subspace containing the $p_z$ orbitals. Thus, 
\begin{equation}
\left[\hat{L}_z\right]_{ij} = \left\langle i \left|\frac{i m_e}{\hbar} \left(  \hat{x}{\cal H}^{GNF}\hat{y}- \hat{y}{\cal H}^{GNF}\hat{x} \right)\right| j \right\rangle \nonumber ~. \\
\end{equation}
Introducing the completeness relation $\hat{I}=\sum_i  \left| i\left\rangle \right\langle i \right|$,
\begin{eqnarray}
    \left[\hat{L}_z\right]_{ij}  & = &
 \frac{i m_e}{\hbar} \sum_{l,k} \left(  X_i\delta_{il}{\cal H}_{lk}^{GNF} Y_{j}\delta_{kj}- Y_i\delta_{il}{\cal H}^{GNF}_{lk} X_j\delta_{kj}\right)  \nonumber \\
& = & \frac{i m_e}{\hbar} \left(  X_i{\cal H}_{ij}^{GNF} Y_{j}- Y_i{\cal H}^{GNF}_{ij} X_j\right) \\ 
 & = & \frac{i m_e t_{ij}}{\hbar} \left(  X_i Y_{j}- Y_i X_j\right). \label{oamij} 
\end{eqnarray}  
This is Eq.~\ref{lz-area} presented in the main manuscript where $\textbf{A}_{ij}\equiv\left(  X_i Y_{j}- Y_i X_j\right)$. We highlight that the structure of the OAM matrix elements is non-zero whenever the Hamiltonian matrix elements are non-zero. Thus, the structure of the OAM is the same as in the Hamiltonian.

\textit{Generalizing to the non-orthogonal case}\label{nonorthogonalLz}---
We can derive the matrix elements of the orbital angular momentum operator for a multi-orbital non-orthogonal basis. To this end, we return to the definition of the angular momentum matrix element between sites $i$ and $j$, now considering a non-orthogonal basis set. We assume the completeness relation \cite{SorianoPalacios2014},
\begin{equation}
 \hat{I} = \sum_{i,j} | i \rangle \left[\hat{\mathcal{S}}^{-1}\right]_{ij} \langle j | ~.
\end{equation}
If we keep only the $p_z$ orbitals, and they are still mutually orthogonal to the remaining atomic orbitals \cite{mulliken_formulas_1949} forming the solution for the isolated carbon atom, 
\begin{equation}
 \langle i | \hat{\mathbf{r}} | j \rangle = \frac{\mathbf{R}_i + \mathbf{R}_j}{2}\mathcal{S}_{ij} ~.
\end{equation}
More generally, we can consider a multi-orbital basis on each site \cite{PhysRevB.64.235111}, {\it i.e.}, $i \rightarrow \left\{i,n,m,l\right\}$. In that case,
\begin{equation}
    \langle i | \hat{\mathbf{r}} | j \rangle = \left(\hat{x}_{ij}, \hat{y}_{ij}\right) ~,
\end{equation}
where inter- and intra-orbital dipole elements are also included.

Either way, in this basis with overlap, $\hat{x}$ and $\hat{y}$ need not commute. Thus we are required to use the symmetrized form of the angular momentum operator resulting in
\begin{widetext}
\begin{eqnarray}
    \left[\hat{L}_z\right]_{ij} & = & \frac{im_e}{2\hbar}\left[\hat{x}\mathcal{S}^{-1}\mathcal{H}^{GNF}\mathcal{S}^{-1}\hat{y} - \hat{x}\mathcal{S}^{-1}\hat{y}\mathcal{S}^{-1}\mathcal{H}^{GNF} - \hat{y}\mathcal{S}^{-1}\mathcal{H}^{GNF}\mathcal{S}^{-1}\hat{x} + \hat{y}\mathcal{S}^{-1}\hat{x}\mathcal{S}^{-1}\mathcal{H}^{GNF} - h.c. \right]_{ij} ~,  \label{aom_nonorth}\\
    & = & \frac{im_e}{\hbar} \sum_{m,n,k,l}  \Big\{   \hat{x}_{im}\left[\mathcal{S}^{-1}\right]_{mn}\mathcal{H}_{nl}^{GNF}\left[\mathcal{S}^{-1}\right]_{lk}\hat{y}_{kj}  -  \hat{x}_{im}\left[\mathcal{S}^{-1}\right]_{mn}\hat{y}_{nl}\left[\mathcal{S}^{-1}\right]_{lk}\mathcal{H}_{kj}^{GNF}     
    \nonumber\\ 
    & - &  \hat{y}_{im} \left[\mathcal{S}^{-1}\right]_{mn}\mathcal{H}_{nl}^{GNF}\left[\mathcal{S}^{-1}\right]_{lk}\hat{x}_{kj}  + \hat{y}_{im}\left[\mathcal{S}^{-1}\right]_{mn}\hat{x}_{nl}\left[\mathcal{S}^{-1}\right]_{lk}\mathcal{H}_{kj}^{GNF} - h.c. \Big\} . \label{aomnonorthfull}
\end{eqnarray}
\end{widetext}
This is the full definition of matrix elements of the angular momentum operator. It also accounts for atomic orbitals carrying intrinsic atomic OAM.  This information will be encoded in matrix elements of the Hamiltonian as well as in the dipole matrix elements. Although much more complicated, we can show that the non-zero OAM is still present, and the operator structure is similar to the simplified case considered in the manuscript (see Supplementary Material file \cite{SM}). 

\end{document}